\documentclass[a4paper,fleqn,usenatbib]{mnras}
\usepackage{txfonts}
\usepackage[T1]{fontenc}
\usepackage{ae,aecompl}

\usepackage{graphicx}	

\usepackage{xcolor,ulem}
\usepackage{cancel}
\usepackage{slashed}
\usepackage{lipsum}
\usepackage{float}
\usepackage{relsize}
\usepackage{multirow}

\def\be{\begin{equation}}
\def\ee{\end{equation}}
\def\bea{\begin{eqnarray}}
\def\eea{\end{eqnarray}}
\def\bn{\begin{enumerate}}
\def\en{\end{enumerate}}
\def\bi{\begin{itemize}}
\def\ei{\end{itemize}}
\def\bfh{\begin{figure}[H]\begin{center}}
\def\efh{\end{center}\end{figure}}

\def\Eiso{E_{iso}}
\def\tv{\theta_v}
\def\tj{\theta_j}
\def\eB{\epsilon_B}
\def\eE{\epsilon_E}
\def\p{p}
\def\Msun{M_\odot}

\def\RAG{\mathbf{R_{\mathsmaller {AG,BNS}}}}
\def\Vdet{\mathbf{V_{\mathsmaller {det}}}}
\def\fgwdet{\mathbf{f_{\mathsmaller {GWdet}}}}
\def\fagdet{\mathbf{f_{\mathsmaller {AGdet}}}}
\def\rBNS{\mathbf{r_{\mathsmaller{BNS}}}}

\def\dg{^\circ}

\graphicspath{{figures/}}

\makeatletter

\newcommand{\Rmnum}[1]{\expandafter\@slowromancap\romannumeral #1@}
\makeatother

\title[Rates of GW-EM joint detections]{Rates of Short-GRB afterglows in association with Binary Neutron Star mergers}
\author[M. Saleem et al.]{
M. Saleem$^{1}$,\thanks{E-mail: saleemc87@iisertvm.ac.in}
Archana Pai$^{2}$,
Kuntal Misra$^{3}$,
L. Resmi$^{4}$,
and K. G. Arun$^{5,6}$
\\
$^{1}$Indian Institute of Science Education and Research Thiruvananthapuram, CET Campus, Trivandrum 659016. \\
$^{2}$Department of Physics, Indian Institute of Technology Bombay, Powai, Mumbai 400076. \\
$^{3}$Aryabhatta Research Institute of Observational Sciences, Nainital. \\
$^{4}$Indian Institute of Space Science and Technology, Trivandrum. \\
$^{5}$Chennai Mathematical Institute, Siruseri, 603103 Tamilnadu.\\
$^{6}$Institute for Gravitation and the Cosmos, Pennsylvania State University, State College, PA 16802.
}

\date{Accepted XXX. Received YYY; in original form ZZZ}

\pubyear{2017}

\begin{document}
\label{firstpage}
\pagerange{\pageref{firstpage}--\pageref{lastpage}}
\maketitle

\begin{abstract}
	Assuming all binary Neutron Star mergers produce Short Gamma Ray Bursts
	(SGRBs), we combine the merger rates of binary Neutron Stars (BNS) from population
	synthesis studies, the sensitivities of advanced Gravitational Wave (GW) interferometer networks, and of the electromagnetic (EM) facilities in various wave bands, to compute the detection rate of associated afterglows in these bands. Using the inclination angle measured from GWs as a proxy for the viewing angle and assuming a uniform distribution of jet opening angle between 3 to 30 degrees, we generate light curves of the counterparts using the open access afterglow hydrodynamics package BoxFit for X-ray, Optical and Radio bands. For different EM detectors we obtain the fraction of EM counterparts detectable in these three bands by imposing appropriate detection thresholds. In association with BNS mergers detected by five (three) detector network of advanced GW interferometers,
	assuming a BNS merger rate of  $0.6-774{\rm Gpc}^{-3}{\rm yr}^{-1}$~\citep{dominik2012double}, we find the afterglow detection rates (per year) to be $0.04-53$ ($0.02-27$), $0.03-36$ ($0.01-19$) and  $0.04-47$ ($0.02-25$) in the X-ray, optical and radio bands respectively. Our rates represent maximum possible detections for the given BNS rate	since we ignore effects of cadence and field of view in EM follow up observations.
\end{abstract}

\begin{keywords}
{gravitational waves -- gamma ray bursts}
\end{keywords}



\section{Introduction}

The observations of {gravitational waves from compact binary merger events during the observational runs of the advanced LIGO detectors (LIGO-Hanford (H) and LIGO-Livingston (L)) and Virgo (V) detector have firmly established the era of gravitational wave (GW) astronomy   \citep{gw150914,gw151226,gw170104,gw170608,gw170814,GW170817}}.
The Japanese detector KAGRA (K) \citep{KAGRA} is under construction and the approved LIGO-India (I) \citep{LIGO-India} detector is expected to come up within a decade. With more number of detectors added, the global interferometric detector network will probe the distant Universe in GWs. 
Binary Black Hole (BBH), Binary Neutron Star (BNS) and Neutron Star-Black Hole (NSBH) mergers are among the prime targets of GW detectors. So far, 13 confirmed BNS systems are known \citep{taursiDNS}. 
NSBH systems have not yet been observed in the electromagnetic (EM) window
and hence the existence of these systems is an open puzzle. However,
various binary formation channels predict their existence and there are
population synthesis models which predict the rates (see for example \citealt{dominik2012double}).

The short gamma ray bursts (SGRBs) with $T_{90}$ duration ( time
over which 90\% of the total observed energy is emitted) less than 2 sec are short intense flashes of gamma rays releasing $\sim 10^{50}$~ergs of energy (see \citealt{BergerReview} for a review). BNS and NSBH mergers (collectively called as
\textit{NS mergers}) are believed to be the plausible
progenitors of SGRBs~\citep{Narayan92}. A few hundreds of SGRBs are detected by all sky scanning $\gamma$-ray instruments. {Recently, the observations of a SGRB (GRB170817A) in spatial and temporal  coincidence with a GW observation from a BNS merger (GW170817)  have given strong direct evidence for the NS-merger progenitor hypothesis \citep{GW170817,MMApaper,GRB+GW-2017}}. The era of multi-messenger astronomy will tremendously improve our understanding of the physical mechanisms of SGRB formation
\citep{arun2014synergy,Bartos2012,granot2017lessons}.

GRBs are believed to be produced from relativistic jets of half opening
angles of the order of a few degrees but due to strong relativistic
beaming, {the chances of detecting} the prompt emission decreases  if the observer's line of sight  {is not oriented to within} the cone of the jet.  The long wave-length counterparts of SGRBs known as the ``afterglows" are
potential  EM counterparts of the GW signal. {The nature of X-ray and radio counterparts of the recent joint event GW170817+GRB170817A was consistent to be as powered from a GRB jet pointed away from our line of sight \citep{GRB+GW-2017,troja2017x,alexander2017electromagnetic,kim2017alma}}. 
{Further, the observation of the optical/UV/NIR counterparts were indicative of the importance of  kilonovae emission as EM counterparts of BNS merger events \citep{smartt2017kilonova,arcavi2017optical,pian2017spectroscopic}}.
There are ongoing efforts for rapid EM follow up of the observed GW events in the EM
window (GW-triggered EM follow-up) \citep{{singer2014EMfollowup},
{kasliwal2014discovering}}. With increasing number of GW detectors,  we
expect the sky error to reduce to as small as a few square
degrees\citep{Fairhurst2010,Klimenko2011,tagoshi2014parameter}.

Coincident GW-SGRB detections have been addressed in literature in numerous
contexts \citep{{Bartos2012},{siellez2013simultaneous}, 
{regimbau2015revisiting}, {clark2015prospects}}. Some of
the previous studies focus on the scenario of joint detectability of
GW transient and SGRBs/afterglows. For example, \cite{metzger2012most}
addresses the detectability issues of the EM-counterparts of the
GW-detected NS mergers. Recently, \cite{lazzati2016off} explored the
detectability of various EM components including prompt emission and afterglow
in different EM bands. However, the authors fixed intrinsic GRB parameters
and distance ($200$~Mpc) but studied the EM counterparts for different
viewing angles. In \cite{ghirlanda2016short}, authors estimate the rate of
coincident SGRB-GW detections from the adLIGO BNS/NSBH detection
volumes using luminosity distribution function and a redshift
distribution of SGRBs. In \cite{patricelli2016prospects} the authors
investigated the detectability of prompt emission and afterglow observed with \textit{Fermi} in coincidence with BNS sources simulated as per population synthesis predictions. They considered three different values of $\tj$ ($0.3, 10,$ and $30 \deg$) in their simulations.
\cite{feng2014detectability} did a rather rigorous study to compute the detectability of late time radio afterglows in low radio frequencies ($10-1800$~MHz), relevant for the Square Kilometer Array (SKA) and its precursors, assuming SGRB rate to be the same as the BNS merger rate obtained in \cite{abadie2010predictions}. They used \textit{BoxFit}, a numerical hydrodynamic broad-band afterglow evolution code, to simulate afterglow light-curve and estimated the rates of radio afterglows \citep{van2012gamma}.

In this work, we estimate the rates of afterglow detections in various bands, associated 
with BNS mergers detected by advanced multi-detector ground based GW
detectors. We consider \textit{Swift} X-ray Telescope (XRT), the Large Synoptic Survey Telescope (LSST) and the Jansky Very Large Array (JVLA) for detections in
X-ray, optical and radio bands respectively(optical afterglow detections with future wide-field instrument \textit{LSST} is compared with the existing wide-field instruments \textit{Pan-STARR1} and \textit{DECam}).  
We only consider the afterglow forward shock emission and calculate the light curves at 
$1$~keV for X-ray, $4.5\times10^{14}$~Hz for optical and $15$~GHz for radio band using \textit{BoxFit}.   We carry out simulations to obtain a synthetic catalogue of mergers of 
BNS detected with 3- and 5-detector networks. As EM counterparts to the synthetic BNS detections, we simulate associated SGRB afterglow light curves, with luminosity distance and the inclination angle of the binary with respect to (w.r.t) the detector
being used as proxies for their association. Apart from distance and
inclination, we distribute other SGRB/afterglow parameters within their
uncertainty regions obtained from the current observations as well as theoretical
understanding of the SGRBs. We explore the afterglow parameter space
rigorously and estimate the joint GW and SGRB afterglow rates for
different distributions of the afterglow parameters.
	\begin{figure*}
		\includegraphics[scale=0.6]{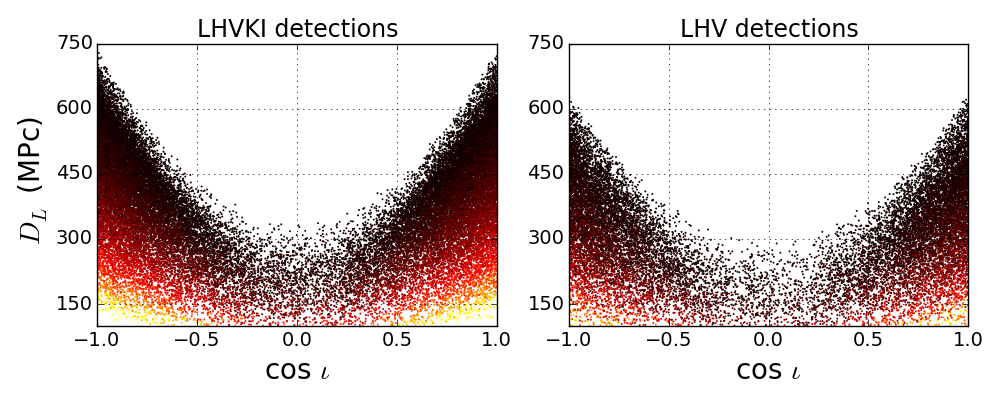}
		\includegraphics[scale=0.42]{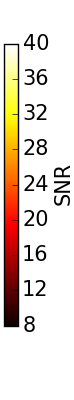}
		\vspace{-3mm}
		\caption{Distance-inclination($D_L-\cos\iota$) scatter plots of the simulated non-spinning BNS sources detected at LHVKI(left) and LHV(right) networks with a detection criterion of minimum network SNR of 8. Face-on sources($\cos\iota\rightarrow\pm 1$) are detected upto much larger distances than edge-on sources ($\cos\iota\rightarrow 0$) whose detections are possible only from nearby distances} 
		\label{BNS-pop}
	\end{figure*}
	
	Our study is based on several assumptions which are listed
        below:
    \bn
	\item BNS merger sources are uniformly distributed in volume
          in the universe with a density as predicted by population
          synthesis models in \cite{dominik2012double}. Here, we
          ignore the effects of redshift since the detectable BNS
          sources are distributed in a closer volume (typically up to
          750 Mpc [See Fig. \ref{BNS-pop}] in which the redshift
          effects can be safely ignored.
	\item We consider all BNS systems to be non-spinning and made
          up of component masses of 1.4$\Msun$ each. 
   	\item We consider that all the BNS sources are observed by the
          LHV network and LHVKI detector network with each of them having noise PSD equal
          to the advanced LIGO designed sensitivity noise. 
    \item We assume that all BNS mergers produce SGRB jets with an associated 
	      long wavelength afterglow in X-ray, optical and radio bands.
    \item The calculations are for a uniform top-hat jet model, 
		  considered in the \textit{BoxFit} code. It is possible that the jet structure could be different.
    \item While generating the light curves using \textit{BoxFit}  code,  no 
	      correction for    extinction (Galactic+host) has been made.
    \item We also ignore the limitations caused by the field of view (FOV) of 
	      EM counterpart follow-up instruments and the cadence of follow-up programs.   In our work we consider that the follow-up observations have a minimum latency of 5 hours in current scenario.
   	\en
	
	The rest of this paper is organized as follows. In
        Section.~\ref{sec-method} we introduce our method of rate
        estimation. In 
        Section.~\ref{sec-gw-sim}, we describe our simulated BNS source
        population and their GW-detectability with a typical 3
        detector network LHV and 5-detector network LHVKI. In
        Section.\ref{sec-ag}, we simulate afterglows as EM
        counterparts to the GW detected BNS sources and estimate their
        detectability with specific X-ray, optical and radio
        instruments. In Section.~\ref{sec-results}, we combine our
        findings and assumptions from previous sections and obtain the
        rates of various coincident joint GW and EM detection
        scenarios. Finally in Section.~\ref{sec-summary}, we summarize our
        work and in Section.~\ref{sec-caveats}, we discuss the caveats and future prospects.

	\section{Method for rate estimation}
	\label{sec-method}
	We use the below expression to calculate the detection rates $\RAG$ of the SGRB afterglows in association with the GW-detected BNS merger events (ie, number of joint GW-afterglow detections per year), 
	\begin{equation}
		\RAG =	\rBNS \times \Vdet \times \fgwdet \times \fagdet ,
		\label{rate-def}
	\end{equation}
	
	where, each quantity introduced in the right hand side of Eq. \ref{rate-def} is described below.
	\bi
	\item $\rBNS$ is the intrinsic BNS merger rate given in units of 
	${\rm Gpc}^{-3} {\rm yr}^{-1}$. 
	In   our case, we use the rates calculated by
	\cite{dominik2012double} based on population synthesis models  which estimates
	the BNS merger rates to be $0.6-774 {\rm Gpc}^{-3} {\rm yr}^{-1}$
	\item $\Vdet$ is the detection volume of the GW detector network defined as the volume 
	of  a sphere centered at earth and extending up to the farthest
source detectable by the given GW detector network  (distance to which is
referred to as horizon of the detector). For 3- and 5- detector networks, we compute this quantity in Section \ref{sec-gw-sim}.
	
	\item $\fgwdet$ is the detection fraction for a given GW detector network, defined as the
	      fraction of BNS mergers detected from a source population which is uniformly distributed within the detection volume $\Vdet$ defined above. Please note that different detector networks have different sky coverage and hence the detection fraction will also vary for them. For  LHV and LHVKI detectors, we compute $\fgwdet$ in Section \ref{sec-gw-sim}.
	
	\item $\fagdet$ is the detection fraction of the afterglow components, defined as 
	      the fraction of BNS-detected sources which has detectable afterglow by EM instruments. With simulated afterglow light-curves,  we compute this quantity for different afterglow components in Section \ref{sec-ag}.
	\ei 
	
	\section{Simulated BNS sources and their GW-detectability}
	\label{sec-gw-sim}
	In this section, we estimate and compare the GW-detection
        volumes of 3- and 5 detector networks different detector
        networks and we also quantify the BNS detectability of each
        network within the detection volume. We consider a 3-detector
        network \textbf{LHV} which includes LIGO-Hanford(H),
        LIGO-Livingstone(L) and Virgo(V), and a 5-detector network
        \textbf{LHVKI} by adding the two upcoming detectors Kagra(K)
        and LIGO-India(I).
	 
	We simulated $3\times10^5$ non-spinning BNS sources with
        component masses 1.4$\Msun$ each, uniformly distributed in
        comoving volume between 100-740Mpc. For each source, we
        simulated GW signal using the analytical 3.5 order
        post-Newtonian waveform \citep{Bliving,BDEI04,BDIWW95}
        and computed the network signal to noise ratio(SNR), defined as the
quadrature sum of the SNRs in the individual detectors~\citep{Pai2000}.
        With a minimum network SNR of 8 as the
        detection criterion, we recover 51834 sources with LHVKI and
        24747 sources with LHV networks. The farthest source detected with
        LHVKI is at $\sim730$ Mpc and for LHV, at $\sim630$ Mpc which
        are defined as the horizon distances of each network and which
        in turn, can be used to compute the detection volumes of the
        networks as described in Section.~\ref{sec-method}. The
        detection fraction $\fgwdet$ is the fraction of detected
        sources normalized w.r.t the total number of sources within
        the detection volume. In other words, detection fraction is
        obtained as the ratio of number of detected sources to the
        total number of sources within the detection volume. Note that
        all the sources within the detection volume are not detected
        because of the directional sensitivity of GW detectors. We
        have listed the detection volumes and detection fractions for
        LHVKI and LHV networks in Table.~\ref{tab-gw-network}.
		\begin{table}
		\centering
		\label{tab-gw-network} 
		\caption { Detection volume and detection fraction in the detection volume for different networks of GW detectors }		
			\begin{tabular}{lccr}
				\hline
				\textbf{Network} & \textbf{Horizon} &
$\Vdet$(${\rm Gpc}^{3}$) & $\fgwdet$  \\
				\hline	
				HLV	    &630Mpc	& 1.05  & 0.133$\pm$0.003 		\\
				HLVKI	&730Mpc	& 1.63  & 0.181$\pm$0.001		\\
				\hline
			\end{tabular}
		\end{table}
		
	Figure.~\ref{BNS-pop} shows the scattered plot of
        distance($D_L$) and inclination angles ($\mathbf{\iota}$) of
        the detected BNS sources. Inclination angle $\mathbf{\iota}$
        is the angle between observer's line of sight and the axis of
        binary orbital plane. Left panel shows the $D_L-\cos\iota$
        scatter plot for the sources detected by the LHVKI network and the
        right panel shows the same for LHV network with the color bar
        showing the SNR. For the actual simulated population which are
        uniformly distributed in volume, distance follows $P(D_L)
        \propto D_L^2$. However, due to the antenna pattern
        functions, the GW detector networks have different
        sensitivities at different directions and hence the distant
        sources will be detectable only if they are located at highly
        sensitive directions. In other words, there will be a drop in
        average SNR for distances close to horizon which
        results in lesser number of detectable sources at larger
        distances~\citep{Schutz2011}.
	
	In the simulated population, inclination $\iota$ of the binary
        is distributed such that $\cos{\iota}$ is uniform between -1 and 1, (ie,
        $P(\cos{\iota}) \propto {\cal{U}}(-1,1)$ which translates as
        $P(\iota) \propto \sin{\iota}$). However, as seen in
        Figure.~\ref{BNS-pop}, the inclination distribution of detected
        sources is biased towards face-on sources
        ($\cos\iota\rightarrow\pm 1$) such that they are detectable to
        much larger distances than the edge-on sources
        ($\cos\iota\rightarrow 0$). Also, the color in
        Figure.~\ref{BNS-pop} shows the SNR distribution of detected
        sources at two networks.
	\section{Simulated Afterglows and their detectability}
	\label{sec-ag}
	In this section, we describe our afterglow simulations associated with the 50,000 GW-detected BNS mergers shown in
        Figure.\ref{BNS-pop}. We have generated afterglow light-curves
        in radio, optical and X-ray bands using the open access
        hydrodynamic simulation package
        BoxFit\citep{van2012gamma}.
	
	\subsection{Associating GWs from BNS mergers  with SGRB afterglows}
	The extrinsic afterglow parameters such as distance, inclination, sky
        location etc can be used  to confirm the association between 
        SGRB afterglow observations and binary neutron star mergers detected
using GWs. Inclination
        $\iota$ of the binary in the GW literature is same as observer's
        viewing angle($\tv$) in the GRB literature,  which is defined as the
angle between the axis of the jet and the observers line of sight.  Here one is
implicitly assuming that the GRB jet is launched along the rotation axis of the
black hole that is formed by the merger of the two neutron stars within the
standard fireball paradigm.
 In our case, we
        use distance $D_L$ and inclination $\iota$(or viewing angle
        $\tv$) as relevant GW inputs for simulating its EM
        counterparts(afterglow for here). Or in other words, using
        $D_L-\iota$ pair as a bridge, we associate a given GW-detected
        BNS merger and a SGRB afterglow to a single physical origin.

	The association would be further strengthened if more 
        parameters can be identified in common for the  GW signal
        and the SGRB afterglow.  The recent works \cite{giacomazzo2012compact},\cite{foucart2012black},\cite{kawaguchi2016}, \cite{dietrich2017modeling} derive  the disk mass $M_{\rm disk}$ and ejecta mass $M_{\rm ej}$ from the intrinsic parameters of progenitor  binary such as component masses, spins and the equation of
        state parameter which are all obtained from GW
        observations. Such burst properties may be further used to derive the isotropic energy $E_{iso}$ of the source. \cite{salafia2017and}  
        considered a similar approach for estimating the 
        detectability of EM counterparts of GW detected binaries 
        and \cite{coughlin2017towards} used the same for exploring the properties of kilonovae lightcurves.  
	    However, given the large uncertainties associated with the estimates of $E_{iso}$ in this method, we have not employed this in the present paper.

	\subsection{Afterglow parameter space and lightcurves}
	Apart from extrinsic  parameters $D_L$ and $\tv$ which are obtained from GW
        detections (it is also possible to considerably refine $D_L$ through
spectroscopy of the optical host-galaxy, if detected), afterglow light curves depend on six intrinsic physical parameters of the emitting plasma. 
        These are the  isotropic equivalent kinetic energy $E_{\rm iso}$ carried by the jet, initial half opening angle of the jet $\tj$,  number density $n$  of the  circumburst
        medium (assumed to be homogeneous), fraction $\eB$ and $\eE$ of the shock thermal energy converted to downstream magnetic field and non-thermal electrons respectively, and power-law index $\p$ of energy-spectrum of the radiating electrons.  

We have simulated 50,000 GRB afterglow lightcurves within the $8$-dimensional
afterglow parameter space described above.  See Table-\ref{tab-param-distr} for
the ranges and distribution of each parameters. In a companion paper \citep{saleem2017agparameterspace}, we describe in detail the justifications for choosing the ranges. The two populations considered differ only in the distribution of $E_{\rm iso}, n,$ and $\eB$. We consider uniform (population-2) and logarithmic (population-1) distributions for these parameters as two extreme examples. 
	
	\begin{table}
		\centering	
		\label{tab-param-distr} 
		\caption {Components of afterglow parameter space along with
			their ranges and distributions.  Prior range and distribution for the
			$D_L-\iota$ combination is obtained from GW-detection criteria. Remaining
			parameters are intrinsic to the afterglow generating mechanism and their prior
			ranges are taken as inferred from current observations. We have considered two
			populations of SGRB sources, namely population-1 and population-2, which differ
			in the distributions of $E_{iso},n$ and $\eB$ parameters. $D_L$ and $\iota$ are directly chosen from the GW-detected population. $\cal{U}$ denotes uniform distribution}
		\begin{tabular}{lccr}
			\hline
			\textbf{Parameter}& \textbf{Range} & \textbf{Population-1} & \textbf{Population-2 }  \\
			\hline	
			$D_L$	       & --          & GW prior	              & GW prior	      \\
			$\tv$	       & --          & GW prior	              & GW prior	      \\
			$\tj$          & $(3\dg,30\dg)$	  & $P(\tj)\propto \cal{U}$ &$P(\tj) \propto \cal{U}$ \\
			$\Eiso$(erg)&$10^{49}-10^{52}$ &$P(\log{\Eiso})\propto \cal{U}$&$P({\Eiso})\propto \cal{U}$ \\ 
			$n$	($cm^{-3})$& 0.0001-0.1  & $P(\log{n})\propto \cal{U}$  & $P(n) \propto \cal{U}$  \\ 
			$\eB$          & 0.01 - 0.1  & $P(\log{\eB})\propto \cal{U}$& $P(\eB)\propto \cal{U}$ \\ 
			$\eE$	       & 0.1         & fixed                  & fixed             \\
			$p$            & 2.5         & fixed	              & fixed  \\												
			\hline
		\end{tabular}
	\end{table}	
	
	Since GW-detected $\tv$ distribution is allowed to have values
        anywhere between 0-180 (though it follows the distribution shown in Fig.\ref{BNS-pop}),
        and $\tj$ is only between 3-30, there are several sources for
        which $\tv>\tj$ (outside-jet). For our population, roughly 85\%
        of the sources are outside-jet. This means that for a large fraction
        of the GW-detected sources, the probability of observing prompt emission is feeble.
        So afterglows become the most interesting EM
        counterparts to NS mergers.
	

        \begin{figure*}
		\includegraphics[scale=0.45]{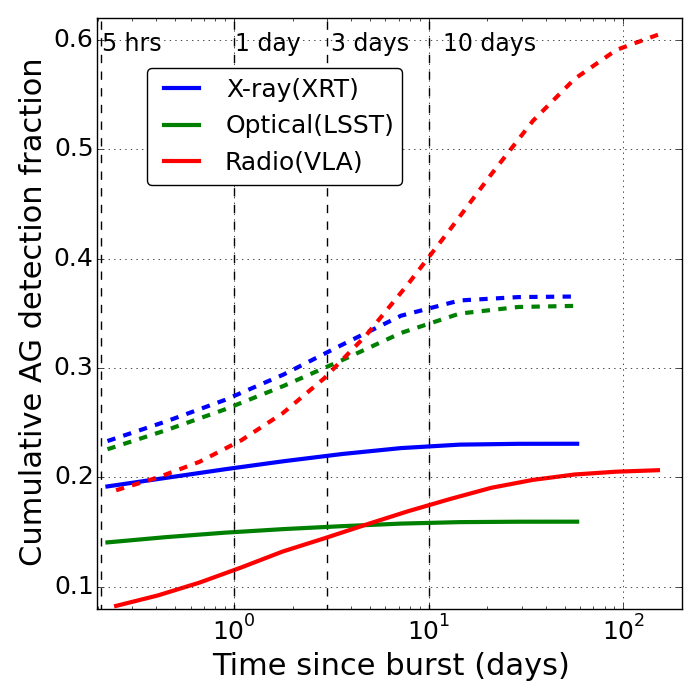}
		\includegraphics[scale=0.45]{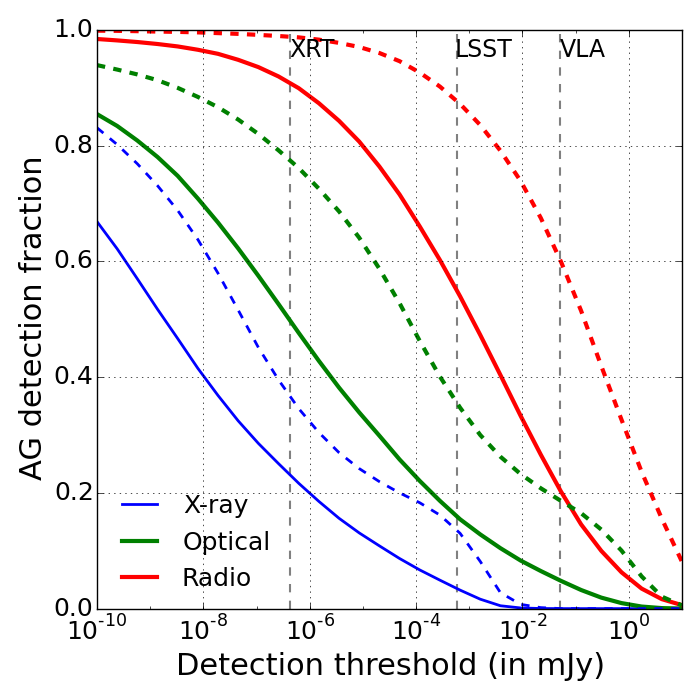}
		\vspace{-3mm}
		\caption{\textbf{[Left]} Cumulative afterglow detection fraction vs time since burst for population-1(solid lines) and population-2(dashed lines) . Figure shows how the detection fraction of different afterglow components grow as a function of the time after the burst for a GW-detected BNS source population with 5-detector network LHVKI. 
		\textbf{[Right]} Detection fraction vs detection threshold. Dashed vertical lines mark the detection thresholds of XRT(X-ray), LSST(optical) and VLA(radio). Figure shows the sensitivity of the detection fractions on detection thresholds}
		\label{fig-EM-detectability}
	\end{figure*}

	\subsection{EM facilities and detections}
	\label{sec-EM-facility}
In this section, we detail the available/future EM facilities for afterglow
detections along with the detection criterion used for this work. The
representative instruments considered for X-ray, optical and radio observations
are \textit{Swift}-XRT, LSST and JVLA respectively. See
Table.~\ref{tab-thresholds} the representative frequencies and threshold fluxes
we used to calculate the rates. We are also demonstrating the effect of
detection sensitivity in the final rates (see next section and Figure.~\ref{fig-EM-detectability} (left)).  

The detection criterion we have followed in all the bands is that a minimum of one point in the light curve should be above the threshold flux. The sensitivity of the instruments depend on the exposure time, nature of the background field, and observation conditions. The latter two factors are arbitrary and hence we are concentrating only on making sure that our detection criteria is consistent with the required exposure time. The \textit{LSST} has a unique combination of wide field capabilities and a sensitive detector reaching up to $\sim 24$~mag in short exposures of $15$~sec.  The quoted \textit{XRT} sensitivity requires an exposure time of $10^4$~sec (ref: \textit{XRT} website). The \textit{JVLA} is capable of achieving $\sim 10 \mu$~Jy rms in $1$~hr of observations. Two adjacent points in our simulated light curves are separated such that $\delta t/t \sim 1$.

We consider 5 hours from the burst as the starting time of EM follow-up observations.  This allows around $2 \times 10^{4}$~sec of $\delta t$. Hence, our detection criterion is consistent with the  exposure time required for achieving the assumed threshold fluxes in all telescopes considered in the study.
However,  in Section.~\ref{sec-cadence}, we are also investigating the effect
of cadence on the final rates. 

	\begin{table}
		\centering
		\caption { Detection thresholds of different instruments}
		\begin{tabular}{lcr}
			\hline
			\textbf{Instrument}\hspace{8mm} & \textbf{Frequency} \hspace{3mm} & \textbf{Flux threshold} \hspace{3mm}\\
			& \textbf{(Hz)} & \textbf{(mJy)}\\
			\hline	
			XRT	    & $2.4 \times 10^{18}$ 	& $4.37 \times 10^{-7}$\\
			LSST	& $4.5 \times 10^{14}$   & $5.75 \times 10^{-4}$ \footnote{Corresponding to a $m_{\rm AB}$ of $24.5$.}\\
			PS1	& $4.5 \times 10^{14}$   & $5.75 \times 10^{-3}$ \footnote{Corresponding to a $m_{\rm AB}$ of $22$.}\\
			DECam	& $4.5 \times 10^{14}$   & $1.2 \times 10^{-3}$ \footnote{Corresponding to a $m_{\rm AB}$ of $23.7$.}\\
			JVLA	    & $1.5 \times 10^{10}$          & $5.0 \times 10^{-2}$\\
			\hline
		\end{tabular}
		\label{tab-thresholds}  
	\end{table}

    \subsection{Afterglow detectability}
    \label{sec-ag-det}
	With the detection criterion described above and the
        instruments and thresholds shown in
        Table.~\ref{tab-thresholds}, we have made detections of
        lightcurves in X-ray, optical and radio bands for both the
        populations. The quantity we are interested is
        the fraction of sources for which we have detectable afterglow,
        denoted as $\fagdet$ as described in Section.~\ref{sec-method}. In
        Figure.~\ref{fig-EM-detectability}, we have shown features of
        detection fractions of different afterglow components.
	  
	The left panel in Figure.~\ref{fig-EM-detectability} shows how
        the detection fraction of different afterglow components grows as a
        function of time after the burst for a GW-detected BNS source
        population with 5-detector network LHVKI. The solid(dashed) curves are
        from SGRB population-1(2). In both the cases, it is observed that
        the X-ray and optical detection fractions reach the maximum in around 10
        days whereas the radio detection fraction takes more than 100 days
        to reach the maximum. In other words, within 10 days, all the detectable
        afterglows are observed in X-ray and optical band whereas radio afterglows may start appearing above the threshold of the instruments several days after the burst.  
        In almost all the observed SGRBs, X-ray and optical afterglow light curves start at
    the peak value and monotonically decreases with time. However, in our
    studies we have seen X-ray and optical afterglow fluxes reach the peak
    within a maximum span of $10$ days. This difference is
        due to the fact that, our study uses the GW trigger and not the GRB
prompt emission and hence our population includes outside-jet sources.
         All the late rising afterglow have come from outside-jet sources,
        whereas the observed SGRBs are all within-jet or close to it. If
        we consider only ``within jet" sources in our simulation, 
        all the detections happens roughly within a day.
	
	For population-1, we detect X-rays for $23\%$($14.5\%$ within-jet +
	$8.5\%$ outside-jet) of the sources, optical for
	$15.9\%$($11.3\%$ within-jet + $4.6\%$ outside-jet) and radio
	$20.6\%$($10.3\%$ within-jet + $10.3\%$ outside-jet) of the
	sources. $14.6\%$ of the sources ($10.2\%$ within-jet + $4.4\%$
	outside-jet) are detected in all 3 bands whereas $73\%$ of the
	sources ($1\%$ within-jet + $72\%$ outside-jet) are not detected in
	any of the bands. For population-2, we detect X-rays for
	36.5\%(15.5\% within-jet + 21\% outside-jet) of the sources, optical
	for 35.7\%(15.5\% within-jet + 20.2\% outside-jet) and radio
	60.5\%(15.5\% within-jet + 4.5\% outside-jet) of the
	sources. $34.5\%$ of the sources ($15.5\%$ within-jet + $19\%$
	outside-jet) are detected in all 3 bands whereas $39.4\%$ of the
	sources ( all outside-jet) are not detected in any of the bands.
			
\begin{table*}
	\centering	
	\caption { Rates of different afterglow components in association with
the BNS mergers detected by 5-detector network LHVKI(3-detector network LHV) for
population-1 and population-2  assuming a BNS merger rate of $0.6-774{\rm
  Gpc}^{-3}{\rm yr}^{-1}$ (and that all BNS mergers produce SGRBs). For optical detection, rates are separately quoted for current wide field instruments \textit{Pan Starrs1} and \textit{DECam} as well as the future wide field instrument \textit{LSST}.
{ The revised BNS merger rate is $320-4740{\rm Gpc}^{-3}{\rm yr}^{-1}$~\citep{GW170817}.
  Consequently, the afterglow detection rates in the table below will also be increased (lower limits by a factor $\sim 530$ and upper limits by a factor $\sim 6$).}}
	\begin{tabular}{lcr}
		\hline
		\textbf{Afterglow observation} & $\RAG$ with population-1  &$\RAG$ with population-2           \\
		& ($yr^{-1}$)                  & ($yr^{-1}$)                \\			
		\hline
		X-ray         & 0.04 - 53  (0.02 - 27) & 0.06 - 83 (0.03 - 42) \\
		Optical(\textit{LSST}) 
					  & 0.03 - 36  (0.01 - 19) & 0.06 - 81  (0.03 - 41) \\
		Optical(\textit{PS1})  
					  & 0.02 - 22  (0.01 - 11) & 0.04 - 57  (0.02 - 28) \\
		Optical(\textit{DECam})
					  & 0.02 - 31  (0.01 - 16) & 0.06 - 72  (0.03 - 36) \\
		Radio         & 0.04 - 47  (0.02 - 25) & 0.11 - 138 (0.05 - 69) \\
		X-ray \&optical(LSST)\&radio  
					  & 0.03 - 33  (0.01 - 18) & 0.06 - 79  (0.03 - 40) \\
		X-ray \&optical(PS1)\&radio    
					  & 0.02 - 21  (0.01 - 11) & 0.04 - 57  (0.02 - 28) \\
		X-ray \&optical(DECam)\&radio  
					  & 0.02 - 30  (0.01 - 16) & 0.06 - 72  (0.03 - 36) \\
		No afterglow  & 0.13 - 167 (0.06 - 77) & 0.07 - 90  (0.03 - 39)  \\
		\hline
	\end{tabular}
	\label{tab-rates} 
\end{table*}

	Though population-2 shows similar trends as in
        population-1, its detection fraction overall increases much more than population-1 in
        particular for radio afterglows. The overall increase in the
        detections can be explained as follows. Due to the uniform
        distribution of $\Eiso$ and $n$, 90\% of the sources have $\Eiso > 10^{51}$
        and $n> 0.01$ in population-2. Whereas, in population-1, where we have uniform
        distributions in $\log{\Eiso}$ and $\log{n}$, only 33\% of the
        sources have $\Eiso > 10^{51}$ and 33\% of the sources have
        $n> 0.01$. This causes majority of the sources in population-2 to
        have relatively higher energy and denser environments
        resulting in an overall increase in the afterglow brightness
        compared to population-1, resulting in more detections. In particular, there is a 
        large increase in radio detections compared to X-ray and optical. The synchrotron peak frequency of the power-law electron spectrum almost always crosses the radio band in the time scale of a few days for all typical ranges of physical parameters. Therefore, the radio light curve reaches the peak in a few days time, by which the jet lorentz factor drops considerably to reduce the effect of relativistic beaming. In addition to that, if there is considerable lateral spreading in the jet, the opening angle also increases by this time. This results in detections of radio afterglows even in cases where X-ray and optical are missed due to viewing angle effects.  Or in other
        words, the highly outside-jet sources can be brought to
        detection in radio band especially if $\Eiso$ and $n$ are high enough whereas the same can not be done for
        X-ray and optical. Precisely
        speaking, among the sources which are detected in radio but
        missed in X-ray and optical for population-2, $\sim 70\%$ of the
        sources have $\tv > 2\tj$, indicating their extremely
        outside-jet scenario.
		
	It should be noted that $66\%(99\%)$ of the within-jet sources
        are detected in all bands whereas $85\%(46\%)$ of the outside-jet
        sources are not detected in any of the bands for
        population-1(2). This clearly indicates that the detectability
        is biased towards within-jet sources. To understand this further,
        we estimated detection fractions in all bands for within/outside jet
        sources separately. Among all the within-jet sources taken
        together, X-ray, optical and radio were detected for
        $94\%(100\%)$, $73\%(99.9\%)$ and $67\%(99.9\%)$ of them
        respectively for population-1(population-2). Conversely, among all the
        outside-jet sources taken together, X-ray and optical were
        detected only for $10\%(24\%)$ and $5\%(24\%)$ of them
        respectively, while radio has been detected for $12\%(53\%)$
        of them for population-1(2).  This means that for X-ray and
        optical detection, what matters is mostly sources' geometric
        profile (within/outside-jet nature) whereas for radio, what matters is
        source' intrinsic properties such as energy and number density
        etc. A more detailed analysis will be presented in a  related
		paper \citep{saleem2017agparameterspace}.
	
		\begin{figure}
			\includegraphics[scale=0.45]{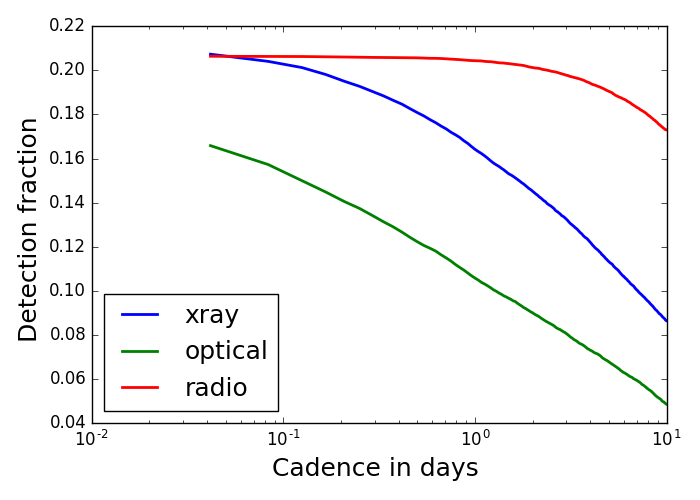}
			\caption{Detection fraction varying as a function of the cadence of the instrument(delay between the first and second observations). The x-axis of the figure shows cadence starting from 1 hour till 10 days and y-axis shows the detection fraction of the afterglow component. It can be observed that the X-ray and optical detection fractions are significantly dropped if a 1-day cadence is assumed instead of 1-hour cadence, whereas radio is unaffected by this. The reason is explained in text.}
			\label{fig-cadence}
		\end{figure}
		
	In the right panel of Figure.~\ref{fig-EM-detectability}, we
        have shown afterglow detection fraction against the detection
        thresholds in $mJy$. The vertical lines mark the
        detection thresholds of \textit{XRT}, \textit{LSST} and
        \textit{JVLA}. Figure shows that all the numbers
        which we compute are highly sensitive to the detection
        thresholds. The bumps observed for X-ray $\sim 10^{-3}$~mJy and optical around $10^{-1}$~mJy in population-2
        are due to the within-jet population. The same results in the saturation observed in population-1 sources at high thresholds. The flux for
        within-jet sources are much higher compared to outside-jet sources due to beaming
        and hence all the within-jet population are detected even with a
        very shallow threshold but the bulk of the outside-jet sources can be
        detected only when the threshold is sufficiently deeper. As
        discussed before, this geometric factor is less of
        a factor for the radio afterglows and the radio curves
        are relatively smoother. The precise location of the bump depends on the ranges of the parameters, especially very strongly on $\tj$, and the type of the distribution (log vs uniform). For example, we can see that due to the higher fraction of high $E_{\rm iso}$ and $n$ sources, population-2 shows the difference predominantly in the form of the bump, while population-1 only shows a saturation. 
		
		\subsection{Detection criterion and detectability}
		\label{sec-cadence}
		As we stated in Section.~\ref{sec-EM-facility}, we have considered one flux point above the threshold as the detection criterion. However, depending on instruments, for a confident detection, just the detection of one flux point may not often be sufficient and observations with a different criterion may affect the detection fraction too. For example, instead of one flux point, if we assume that at least two flux points are required for a confident detection, it requires that after the first observation, the telescope will have to be further scheduled for a second one with a certain time delay between them (known as \textit{cadence}). Different instruments can have different \textit{cadence}. In that case, what metters for the detectability is whether or not the  afterglow flux lasts above the threshold for a duration at least equal to the delay between first and second observations. Consequently, the afterglow lightcurve whose peak flux is above the threshold are likely to be not detected if the flux does not remain above the threshold for a duration equal to the cadence of the instrument. This causes a drop down in the detection fraction. In this section, we demonstrate how our detection fractions in Section.~\ref{sec-ag-det} would have got affected if we were to consider two flux points above the threshold as the detection criterion. 
		
		Figure-\ref{fig-cadence} shows the variation in detection fractions as a function of cadence. The x-axis of the figure shows cadence starting from 1 hour till 10 days and y-axis shows the fraction of lightcurves whose flux lasts above the threshold for a duration at least equal to the cadence(Note that those which lasts for a shorter duration will not be detected). It can be observed that the X-ray and optical detection fractions are significantly dropped if a 1-day cadence is assumed instead of 1-hour cadence, whereas radio is unaffected for this. This is because most of the X-ray and optical detection are from within-jet sources and which are at the peak right from the beginning of observation with a rapidly decaying profile and hence falls below the threshold immediately whereas the radio lightcurves which rises relatively later, lasts for longer duration and hence a 1-day cadence will not affect them. Moreover, a considerable fraction of radio detections comes from outside-jet sources which peaks much later and decays slowly. Due to this, detectable radio afterglows are likely to stay much longer than X-ray and optical. 
		    
	\section{Rates}
	\label{sec-results}
	In this section, we compute the detection rates of afterglows in
        association with the GW-detected BNS mergers(see
        Eq.\ref{rate-def}) by combining the detection fractions and
        detection volumes computed in previous sections(see
        Table.~\ref{tab-gw-network} and
        Figure.~\ref{fig-EM-detectability}) with the BNS merger rates
        obtained from population synthesis models. We adopt the
        numbers obtained in \cite{dominik2012double} which estimates
        the BNS merger rates to be  
        $\rBNS = 0.6-774 {\rm Gpc}^{-3} {\rm yr}^{-1}$
	
	With the EM facilities listed in  Table.~\ref{tab-thresholds} and with
        the GW networks LHVKI and LHV, we have shown our estimates of
        coincident GW(BNS)-afterglow detection rates in
        Table.~\ref{tab-rates}. Estimates with both population-1 and population-2
        are shown in separate columns and numbers are quoted as their
        uncertainty ranges rather that mean or median. The numbers
        outside the bracket are the afterglow detection rates in association
        with the BNS merger detection by the future 5-detector network
        LHVKI and the numbers inside the bracket are for the existing
        3-detector network LHV.
	
	 The dominant source of uncertainty in the estimates (lower-upper bounds of
        rates) arise from the assumptions which go into the population synthesis
        models whose predictions for BNS merger rate we crucially employ in our
		analysis. The other ingredients of rates such as detection
        fractions, detection volume would also have uncertainties in
        their estimates. GW detection fraction was estimated from a
        simulation which was repeated for 100 trials and found that
        the 1-sigma deviation is only less than 1\% and hence we have not
        included those uncertainties. detection volume is the volume
        enclosing all the detections from all these 100 trials(Each
        trial essentially has $3\times 10^{5}$ sources).
    \section{Caveats}
	\label{sec-caveats}
		This paper estimates the joint observation rates of GW BNS merger events and the SGRB afterglow events in the X-ray, optical and radio band. For such a
	    scenario, we have made several assumptions which we revisit below and state possible caveats.
    
	    \subsection{GW-detectability}
	    	Among the ingredients of joint detection rates in Eq.\ref{rate-def}, the GW detection fraction $\fgwdet$ and detection volume $\Vdet$ have been estimated from our simulated BNS sources, using a detection criterion of network SNR of 8. This might be an optimistic criterion compared to realistic cases where it may require to satisfy minimum SNRs at individual detectors. Using a different criterion will alter the detection fraction and detection volumes. Further, we have computed SNR using the designed sensitivity (PSD) curves of advanced LIGO. However, in real cases, as discussed in \cite{nissanke2010exploring}, the particular noise realisation of a particular source will introduce errors in the SNR and hence sub-threshold events can go above the threshold and vice versa, 
	    	affecting the detection fractions slightly. 
	    		
	    	We have fixed the NS mass at $1.4 M_{\odot}$. The BNS detectability can have minor variations considering the range of NS masses as $1-3 M_\odot$.
	    	Further, NSBH are also plausible sources for SGRB. We did not
	    	consider this in our study. 
	    		    		
	    \subsection{Afterglow detectability}
		    Another major ingredient of the joint detection rates in
Eq.\ref{rate-def} is the afterglow detection fraction $\fagdet$ which we have
estimated in Section.\ref{sec-ag} and shown its features in Figure.~\ref{fig-EM-detectability}. 
		    $\fagdet$ estimation is based on several assumptions. First
of all, in each band, we have used specific instruments as well as detection
limits as given in Table.~\ref{tab-thresholds}. These detection limits are subject to changes for practical and observational reasons and hence to affect the estimates of $\fagdet$.
		        
		    In the current study, we have assumed that all the observations start at $t_{start}=5$ hours since the burst(where we also assume a temporal coincidence between the SGRB burst and the BNS merger). However this is subject to the  latency of GW-trigger circulation(see \cite{nissanke2013identifying} for a detailed discussion). In the current scenario, more than 5 hour latency to start EM observations is not unlikely and this can cause slight decrease in $\fagdet$. In a future scenario where we may have automated follow-up facilities, it may be possible to start EM observation with a latency of as small as 1 hour or even earlier. We have found that if the observation starts at 1 hour instead of 5 hours, we get $\sim 10$\% more detections in X-ray,and optical with population-1 whereas no improvement with population-2. Since radio comes later, it is mostly unaffected by the choice of $t_{start}$, but likely to be affected by end time $t_{end}$ of observations. We have used  $t_{end}=\sim 150$ days for radio. However, radio flux can stay much longer and some of the late time radio afterglows can peak after this epoch. So, this earlier $t_{end}$ would also have caused a slight reduction in the rate of radio afterglows. See \cite{feng2014detectability} for a detailed discussion of late time radio afterglows.   
	    
		    As we stated in the beginning, we have assumed sufficient EM facilities to make sure  100\% follow-up to GW triggers. However, in reality, in many cases, the error regions of sky localizations provided by GW observation will be pretty large(several sq. degrees) and it will be often difficult to scan such large error regions and hence to detect the afterglows. 	    
		    Further, in a GW-triggered EM follow-up where the prompt emission is missed, for a large fraction of the outside-jet sources, only radio afterglows will be detectable and that too after several days. In such cases, there will be practical limitations to associate them to the GW event. Late arrival and large error regions from GW observations make it difficult to establish a temporal as well as spatial coincidence. However, such sources also have contributed to our afterglow detections  and therefore we have got optimistic rate estimates. 
	    	
	    \subsection{Parameter ranges and distributions}
	        We have estimated joint detection rates using two populations of
SGRB sources (1\&2). First of all, our knowledge about the universal
distribution of SGRB parameters is very limited. Unlike what we have chosen, the
actual population could be a combination of population-1 and population-2 or
even other populations. This could have affected our rate estimates. Now, even
if we assume that population-1 (or population-2) is the universal distribution
of SGRB sources, for a particular class of progenitor type(BNS in our case), the
distribution of the intrinsic SGRB parameters such as $\Eiso,\tj$ might be
different from what we considered. For eg, $\tj$ for BNS progenitors may be
different from those of NSBH. Or in other words, the ranges we considered for
$\Eiso, \tj$ etc in our study may have perhaps gone beyond the realistic ranges
allowed for BNS progenitors since what we have used is ranges as inferred from
observations which can have various progenitor types. \cite{hotokezaka2016radio} has used different energy values based on progenitor types where BNS and NSBH gives different range of $E_{iso}$ values. Such effects could have biased our rate estimates. 
	    	
	\section{Conclusions and future prospects}
    \label{sec-summary}
	
	In this work, we have targeted to estimate the rate of
        detections of SGRB afterglows in association with the BNS
        merger events detected by GW detector networks.  To compute
        rates, we have combined BNS merger rates given by population
        synthesis predictions in \cite{dominik2012double} with the
        detectability of GW detector networks and the afterglow
        detectability of EM instruments as described in
        Section.\ref{sec-method}.
         
    The GW detectability (or detection fractions) is computed from a
    simulation of $3 \times 10^5$ BNS mergers uniformly distributed in
    comoving constant volume and using network SNR of 8 as the
    detection criterion. This gives a catalogue of BNS detection
    events for network configurations of 3 and 5 detectors. We noted
    the respective detection volume and detection fractions as given
    in Table.~\ref{tab-gw-network}. To each (detected) BNS merger in
    this catalogue, we associated an SGRB source using $D_L$ and
    $\tv$(or inclination $\iota$) as parameters for the GW-EM
    association. Amongst the rest of the SGRB parameters, $E_{iso},n$
    and $\eB$ are drawn from two types of distributions whose details
    are shown in Table.~\ref{tab-param-distr}, namely population-1 and
    population-2.  In order to assess the afterglow detectability, for
    all the sources in both the populations, we simulated afterglow
    lightcurves in X-ray, optical and radio bands using the \textit{BoxFit}
    package and detection fractions are estimated using detection thresholds
    of specific EM instruments namely {\it Swift}-XRT(for X-ray), current wide field optical telescopes \textit{Pan-STARRS1}, \textit{DECam} and the future wide field optical telescopes LSST  and the radio instrument JVLA. The details of instruments including their
    detection thresholds are shown in Table.~\ref{tab-thresholds}.
    
    The afterglow detection fraction has been shown as a function of
    the time after burst in Figure.~\ref{fig-EM-detectability}. We have
    seen that within 10 days, all the detectable afterglows are observed in
    X-ray and optical band whereas radio afterglows may start
    appearing in the instruments several days after the burst. In almost
    all the observed SGRBs, X-ray and optical afterglow light curves start at
    the peak value and monotonically decreases with time. However, in our
    studies we have seen X-ray and optical afterglow flux reaches the peak
    within the maximum span of ten days. This difference
    is due to the fact that our population includes outside-jet
    sources($\sim 85\%$ sources of populations) and all the late rising afterglow have come from outside-jet sources, whereas the observed SGRBs are all
    within-jet sources.  Though population-1 and population-2 shows
    similar trends (Figure.\ref{fig-EM-detectability}), the detection fraction is higher for
    population-2 for all 3 bands in general, and much higher for radio
    in particular. This is because majority of the sources in
    population-2 have relatively higher energy and denser environments
    compared to population-1 and this results in an overall rise in
    the afterglow brightness and hence in the detection fractions.  The
    large increase in radio detections is because, even highly
    outside-jet faint sources can be brought to detection in radio band
    by increasing $\Eiso$ and $n$ whereas the same can not be done 
    for X-ray and optical. This is due to the relatively lower relativistic beaming 
    of radio emissions compared to X-ray and optical emissions. To summarise, what matters for
    X-ray and optical detection is mostly whether the source is
    within-jet or outside-jet whereas what matters for radio detection is
    mostly the energy and number density.
	
    Combining above estimates, we have computed joint GW-afterglow detection
rates for both population-1 and 2. The numbers have been shown in Table.~\ref{tab-rates}.
{It is to be noted that the BNS merger rates estimated from the first two observation runs of LIGO-Virgo is several factors larger than the one which is used in this paper. The revised BNS merger rate is $320-4740{\rm Gpc}^{-3}{\rm yr}^{-1}$~\citep{GW170817} whereas the rates used in this paper is $0.6-774{\rm Gpc}^{-3}{\rm yr}^{-1}$~\citep{dominik2012double}. Consequently, the afterglow detection rates in Table. \ref{tab-rates} will also be increased (lower limits by a factor $\sim 530$ and upper limits by a factor $\sim 6$). This will make the joint scenario of GW and AG detection more frequent in the advanced multi-detector era.}

	\section{Acknowledgements}
	MS thanks the Max Planck Partner Group HPC facilty at IISER-TVM
	and the HPC facility at IUCAA Pune
        where most of the numerical exercise is carried out. Development of the
Boxfit code was supported in part by NASA through grant NNX10AF62G issued
through the Astrophysics Theory Program and by the NSF through grant
AST-1009863. Simulations for BOXFIT version 2 have been carried out in part on
the computing facilities of the Computational Center for Particle and
Astrophysics (C2PAP) of the research cooperation "Excellence Cluster Universe"
in Garching, Germany. KGA is partially  supported by a grant
from Infosys Foundation. AP thanks the IIT-SEED grant for the research support. We would also like to thank Michael Coughlin for critical review as well as insightful inputs which helped to improve the clarity of the manuscript.This manuscript has  LIGO document number {\tt P1700248}.
	\bibliographystyle{mnras}

	\bibliography{draft}

	\end{document}